\documentstyle[11pt,aaspp4]{article}

\begin{document}

\title{
	The Caltech Faint Galaxy Redshift Survey XII:\ 
	Clustering of Galaxies\altaffilmark{1}
}
\author{
	David W. Hogg\altaffilmark{2,3},
	Judith G. Cohen\altaffilmark{4}, and
	Roger Blandford\altaffilmark{5}
}
\altaffiltext{1}{
Based on observations made at the W. M. Keck Observatory, which is
operated jointly by the California Institute of Technology and the
University of California; and at the Palomar Observatory, which is
owned and operated by the California Institute of Technology.}
\altaffiltext{2}{
Institute for Advanced Study, Olden Lane, Princeton NJ 08540, USA;
{\tt hogg@ias.edu}}
\altaffiltext{3}{
Hubble Fellow}
\altaffiltext{4}{
Palomar Observatory, California Institute of Technology,
Pasadena CA 91125, USA}
\altaffiltext{5}{
Theoretical Astrophysics, California Institute of Technology,
Pasadena CA 91125, USA}

\begin{abstract}
A clustering analysis is performed on two samples of $\sim 600$ faint
galaxies each, in two widely separated regions of the sky, including
the Hubble Deep Field.  One of the survey regions is configured so
that some galaxy pairs span angular separations of up to $1$~deg.  The
median redshift is $z_{\rm med}\approx 0.55$.  Strong clustering is
obvious, with every pencil-beam field containing a handful of narrow
redshift-space features, corresponding to galaxy structures with sizes
of 5 to 20~Mpc.  The structures are not obviously organized on planes,
though one prominent, colinear triplet of structures is observed,
spanning $\sim 20$~Mpc.  This may be evidence of a filament.  A
galaxy--galaxy correlation function calculation is performed.  No
significant evolution of clustering (relative to stable clustering) is
found in the redshift range $0.3<z<1.0$.  This is not surprising,
since uncertainties in the correlation amplitude estimated from
surveys like these are large; field-to-field variations and
covariances between data points are both shown to be significant.
Consistent with other studies in this redshift range, the
galaxy--galaxy correlation length is found to be somewhat smaller than
that predicted from local measurements and an assumption of no
evolution.  Galaxies with absorption-line-dominated spectra show much
stronger clustering at distances of $<2$~Mpc than typical field
galaxies.  There is some evidence for weaker clustering at
intermediate redshift than at low redshift, when the results presented
here are compared with surveys of the local Universe.  In subsets of
the data, the measured pairwise velocity dispersion of galaxies ranges
from $200$ to $600~{\rm km\,s^{-1}}$, depending on the properties of
the dominant redshift structures in each subset.
\end{abstract}


\section{Introduction}

Numerical simulations of the growth of large scale structure (LSS) in
the Universe predict that galaxies at the present day ought to lie on
sheets or filaments of thickness a few Mpc, separated by distances of
tens to hundreds of Mpc, and that these structures have been forming
from early times right up to the present day (eg, Buryak et al 1994;
Bond et al 1996; Colberg et al 1999).  As in the numerical models
(although perhaps less clearly), galaxies in the local Universe are
indeed observed to populate such structures (eg, J\^oeveer et al 1978;
Geller \& Huchra 1989; da~Costa et al 1994; Shectman et al 1996;
Vettolani et al 1997).  However, the evolution of these structures
with cosmic time has not been established empirically in more than a
limited way.

In principle, almost any observable aspect of LSS evolution is a
strong function of cosmological parameters, including the density and
age of the Universe, the nature of the dark matter, and the spectrum
of initial perturbations.  In accordance with the general principle
that all cosmological tests are much more difficult than they at first
appear, the observational situation has turned out to be
disappointing, since almost any apparent evolutionary behavior,
especially on small scales, is explained not just as an evolution of
the LSS itself but is combined with an evolving relationship between
mass and light (eg, Baugh et al 1999; Pearce et al 1999), which is to
a large extent unconstrained, both theoretically and observationally.
Even measurements of clustering with a time baseline out to redshifts
$z>3$, which have a huge cosmological ``lever arm'' are thought to
place stronger constraints on the bias than the growth of structure
(Steidel et al 1998; Giavalisco et al 1998; Adelberger et al 1998)!

Since theoretical predictions are turning out to be soft, the Caltech
Faint Galaxy Redshift Survey (CFGRS) has the goal of contributing to
an empirically based history of galaxies in the Universe, which will
constrain detailed theories.  In what follows, we use a pair of
redshift samples, containing over 1000 galaxies to redshift unity, and
spanning a range of angular separations from arcseconds to degrees, to
constrain the sizes, abundances, morphologies, masses, and evolution
of galaxy structures as a function of cosmic time from when the
Universe was roughly half its present age to the current epoch.

After the catalogs of galaxy spectra and photometry are described in
Section~\ref{sec:catalogs}, a traditional galaxy--galaxy correlation
function is measured in Section~\ref{sec:transverse}.  The pairwise
velocity dispersion of galaxies is investigated in
Section~\ref{sec:pairwise}.  Highly significant groups of galaxies are
identified in one of our survey regions in Section~\ref{sec:groups}.
The sizes and morphologies of these structures are investigated in
Section~\ref{sec:groupgal}.  Our conclusions are summarized in
Section~\ref{sec:summary}.

Unless mentioned otherwise, the adopted world model is $H=60~{\rm
km\,s^{-1}\,Mpc^{-1}}$ (or $h=0.6$), $\Omega_M=0.3$, and
$\Omega_{\Lambda}=0.0$.  All magnitudes are Vega-relative.

\section{The catalogs}
\label{sec:catalogs}

This study makes use of large faint galaxy redshift samples from the
CFGRS, in two regions of the sky, one centered on
00\,53\,23~12\,33\,58 (J2000), known as the ``J0053+1234 region'' the
other centered on 12\,36\,49~62\,12\,58 (J2000), known as the ``HDF
region''.  The HDF region is centered on the Hubble Deep Field image
taken with the Hubble Space Telescope (Williams et al 1996).

\subsection{The J0053+1234 region}

In a previous paper (Cohen et al 1999a) redshifts were presented from
a survey in a $2.0\times 7.3~{\rm arcmin^2}$ field, J0053+1234.
Redshifts were obtained for 139 galaxies and identifications were made
of 24 Galactic stars in a flux-limited sample of 195 sources to
$2.2~{\rm \mu m}$ flux limit $K\leq 20~{\rm mag}$.  The sample is
84~percent complete at $K\leq 20$~mag.  The redshifts go to $z=1.44$,
with the median extragalactic redshift $z=0.58$.  This sample is, by
itself, inadequate for study of galaxy clustering.  The size of the
field is determined primarily by the properties of the Low Resolution
Imaging Spectrograph (LRIS; Oke et al 1995) at the Keck Observatory,
used for all of our spectroscopic observations; the field corresponds
to only $\sim 1\times 3~{\rm Mpc^2}$ (proper) in our cosmology at
typical redshifts of roughly one half.

\begin{figure}
\plotone{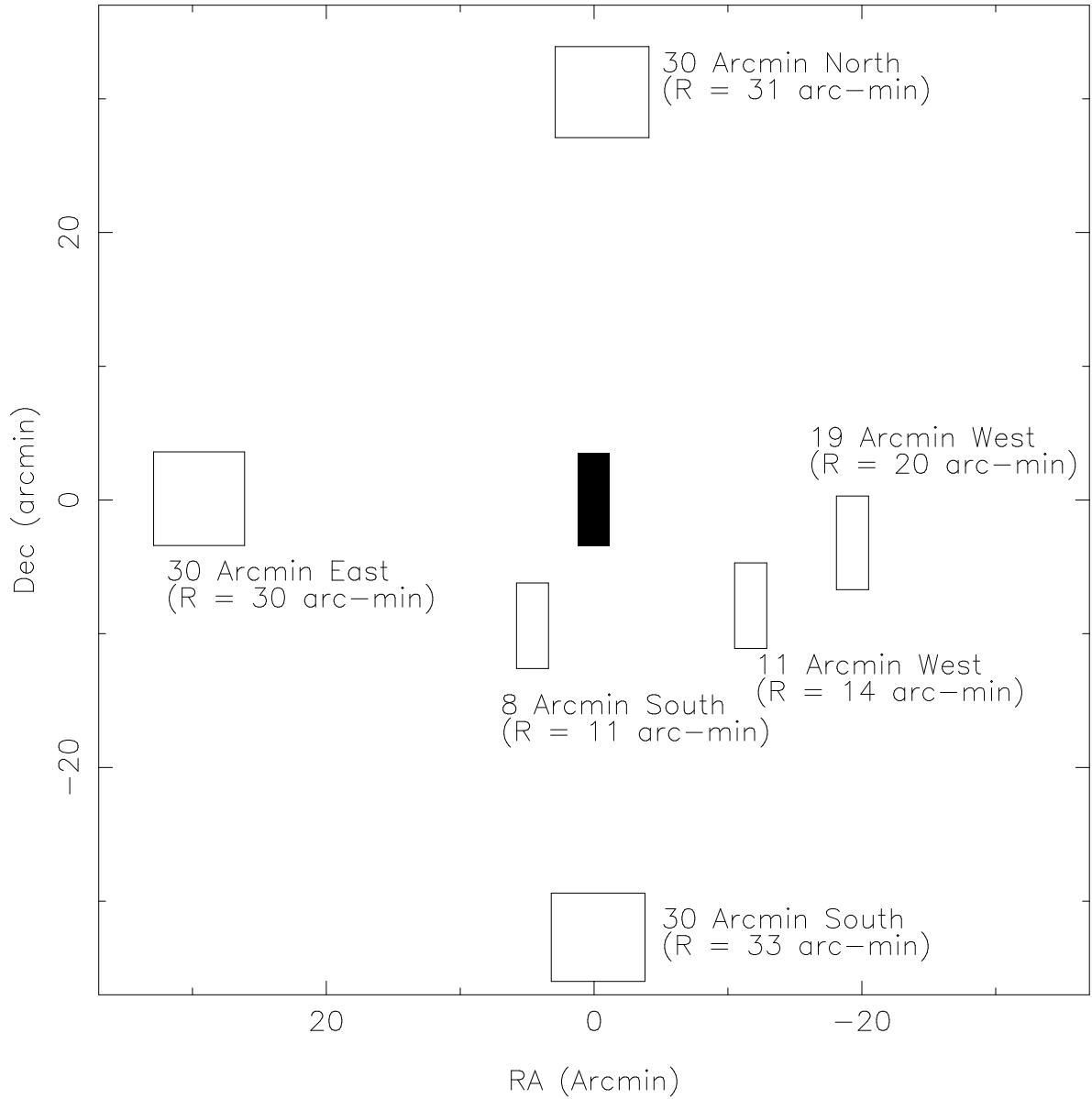}
\caption[]{The layout of subfields in the 0053+1243 region.}
\label{fig:layout}
\end{figure}

To study spatial baselines of 10 to 30~Mpc, it is necessary to span
angular extents of 1~deg.  Given limits on available observing time,
it is impossible at present to obtain a complete redshift sample to
this depth over a contiguous solid angle of this diameter.  In the
0053+1234 region, as will be described in more detail elsewhere (Cohen
et al in preparation), six additional patches of sky (``subfields'')
were observed, distributed over a $1.2\times 1.2~{\rm deg^2}$ area on
the sky, surrounding the main J0053+1234 subfield.  The configuration
of the subfields is shown in Figure~\ref{fig:layout}.  The displayed
area of each subfield corresponds to the area covered by
spectroscopically observed sources.  Table~\ref{tab:fields} gives
various properties of the subfields, including the offsets of their
centers from the center of the main subfield.

\begin{figure}
\plotone{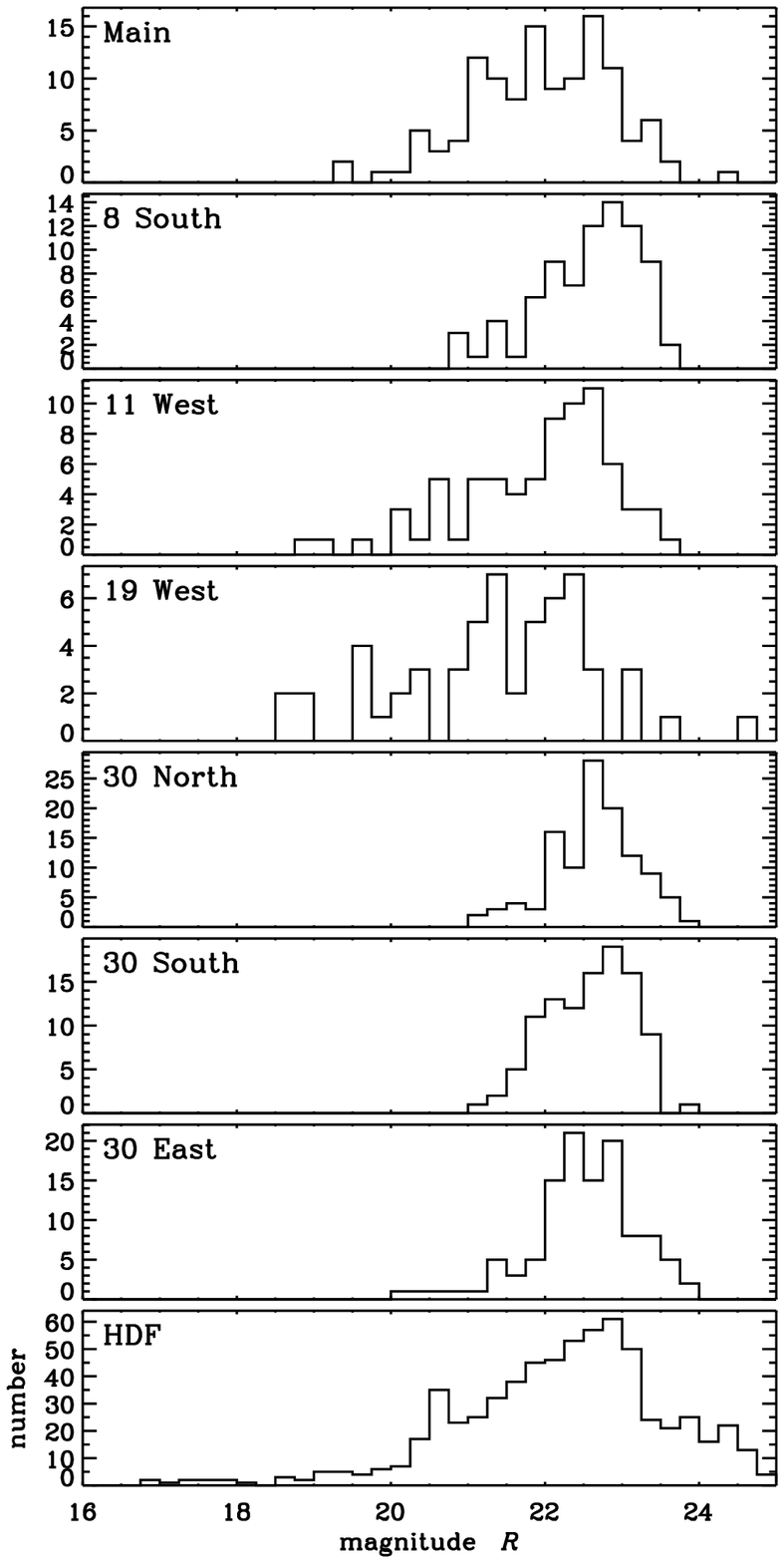}
\caption[]{The numbers of sources as a function of $R$-band magnitude
in the subfields of the J0053+1234 region and in the HDF region.  Note
that different panels have different vertical scales.  Note the
avoidance of bright sources in the subfields other than Main.}
\label{fig:maghist}
\end{figure}

For efficiency, the selection of sources (performed with LRIS images
taken in the $R$ band) was not designed to produce complete samples in
all of the subfields.  The $R$-band photometry consists of 3-arcsec
diameter focal-plane aperture magnitudes.  Bright sources with
$R<22$~mag were eliminated to reduce the fraction of low redshift
sources and thereby reduce telescope time involved in obtaining a
significant sample with $z\sim 0.5$.  Faint sources with $R>23.5$~mag
were also removed, as the probability of successful spectroscopic
identification in only 2 hours of integration is not high.  In
addition, obviously stellar sources with $R<22.5$~mag were avoided,
although a few were observed nonetheless given the vagaries of slitmak
design and the need for bright setup sources with which to align the
masks.  Furthermore, even within the restricted magnitude range
$22\leq R\leq 23.5$~mag, not all sources were observed.  However, a
large enough number of slitmasks were used in each field to wash out
any spatial structure in the selection function.
Figure~\ref{fig:maghist} shows the histogram of $R$-band magnitudes
for all galaxies in each subfield.  Although the selection function is
complicated, it is entirely based on 3-arcsec diameter $R$-band
magnitudes; in principle all important information about the selection
function is therefore contained in Figure~\ref{fig:maghist}.

The spectroscopic observations have a resolution corresponding to
$\approx100~{\rm km\,s^{-1}}$.  The redshifts and galaxy spectral
classes in the new subfields were determined as in the main subfield
(described in Cohen et al 1999a).  The numbers of galaxies with
measured redshifts and other field information are given in
Table~\ref{tab:fields}.  The redshift completeness within the sample
of sources observed exceeds 90~percent in all the subfields.  The
total sample of galaxies in the 0053+1234 region with known redshifts
is 729, with median redshift $z_{\rm med}=0.55$.
Figure~\ref{fig:zhist} shows redshift histograms for the subfields.
The J0053+1234 sample is comparable in size to the entire CFRS survey
(Crampton et al 1995).

\subsection{The Hubble Deep Field region}

The CFGRS also has obtained and compiled 610 galaxy redshifts in a sky
region centered on the HDF.  The sources in this region are selected
in the $R$ band.  The sample is 92~percent complete to $R=24$~mag in
the deep HST-imaged portion of the HDF region and 92~percent complete
to $R=23$~mag in a circular field of $8$~arcmin diameter centered on
the deep portion.  The imaging data, source selection procedures, and
redshift catalog are described elsewhere (Hogg et al 2000; Cohen et al
2000).  Figures~\ref{fig:maghist} and \ref{fig:zhist} show the
histograms of $R$-band magnitudes and redshifts for all galaxies in
the HDF region.

\subsection{Random catalogs}

The studies of galaxy clustering presented below require comparison
with a random or Monte-Carlo sample with little or no clustering but
identical selection criteria, in terms of sky position, magnitude and
redshift.

A representative redshift distribution was constructed for the
galaxies in each one-magnitude-wide bin in $R$ magnitude by smoothing
the observed redshift distribution of galaxies in the redshift bin
with a gaussian of $\sigma=3\times 10^4~{\rm km\,s^{-1}}$ in
rest-frame velocity.  Random catalogs were created for each region by
assigning to each real galaxy position 100 new redshifts chosen from
the smoothed redshift distribution constructed for galaxies in that
real galaxy's magnitude bin.  This procedure results in samples with
good approximations to the angular and radial selection functions of
the true samples, including the complication that the source selection
criteria are different in the different fields.  Note that the random
catalog is 100 times the size of the real catalog.

Because there is a small but significant angular clustering of faint
galaxies, the random sample does not have strictly vanishing spatial
correlations.  However, the angular correlation length of faint
galaxies to $R\approx 24$ is $\approx 0.3$~arcsec (Brainerd et al
1995).  Typical galaxy--galaxy separations in the sample are on the
order of a few Mpc, corresponding to hundreds of arcsec at the typical
redshifts, so the bias introduced by this non-vanishing angular
correlation is very small.

\section{Galaxy--galaxy correlation function}
\label{sec:transverse}

The standard statistic for galaxy clustering is the galaxy--galaxy
correlation function.  The three-space correlation function $\xi(r)$
is not directly computed here, because there are significant
redshift-space distortions.  Rather, a projected correlation function
$\omega(R_{\perp})$ is found through the angular correlation of
galaxies close in redshift.  For each galaxy in the real catalog
(signified by ``D'' for ``data''), the number of other galaxies in the
D catalog, within $\Delta v_r=1000~{\rm km\,s^{-1}}$ in rest-frame
radial velocity, is found in a set of perpendicular proper distance
$R_{\perp}$ bins to make the number of data-data (DD) pairs.  For each
galaxy in the D catalog, the number of galaxies in the random (``R'')
catalog within $\Delta v_r$ is found in the same set of $R_{\perp}$
bins (and divided by 100) to make the DR pairs.  For each galaxy in
the R catalog, the number of galaxies in the R catalog within $\Delta
v_r$ is found (and divided by $10^4$) to make the RR pairs.  The
correlation function estimator is
\begin{equation}
\omega(R_{\perp})= \frac{DD-2\,DR+RR}{RR}
\end{equation}
(Landy \& Szalay 1993).

Only galaxies in the redshift range $0.10<z<1.15$, where the CFGRS is
expected to be substantially complete (Hogg et al 1998), were used in
the correlation function estimation.

This correlation function estimate is to be compared with a projection
of the three-space correlation function $\xi(r)$ of the form
\begin{equation}
\omega(R_{\perp})= \int_{-\Delta v_r}^{\Delta v_r}\,
\xi\left(\sqrt{r^2+\frac{v_r^2}{H(z)^2}}\right)\,\frac{dv_r}{2\,\Delta v_r}
\label{eq:project}
\end{equation}
where $\Delta v_r$ is the $1000~{\rm km\,s^{-1}}$ velocity width used
in the estimation, and $H(z)$ is the Hubble constant as a function of
redshift
\begin{equation}
H(z)= H_0\,\sqrt{\Omega_M\,(1+z)^3+\Omega_k\,(1+z)^2+\Omega_{\Lambda}}
\end{equation}

\subsection{Results}

\begin{figure}
\plotone{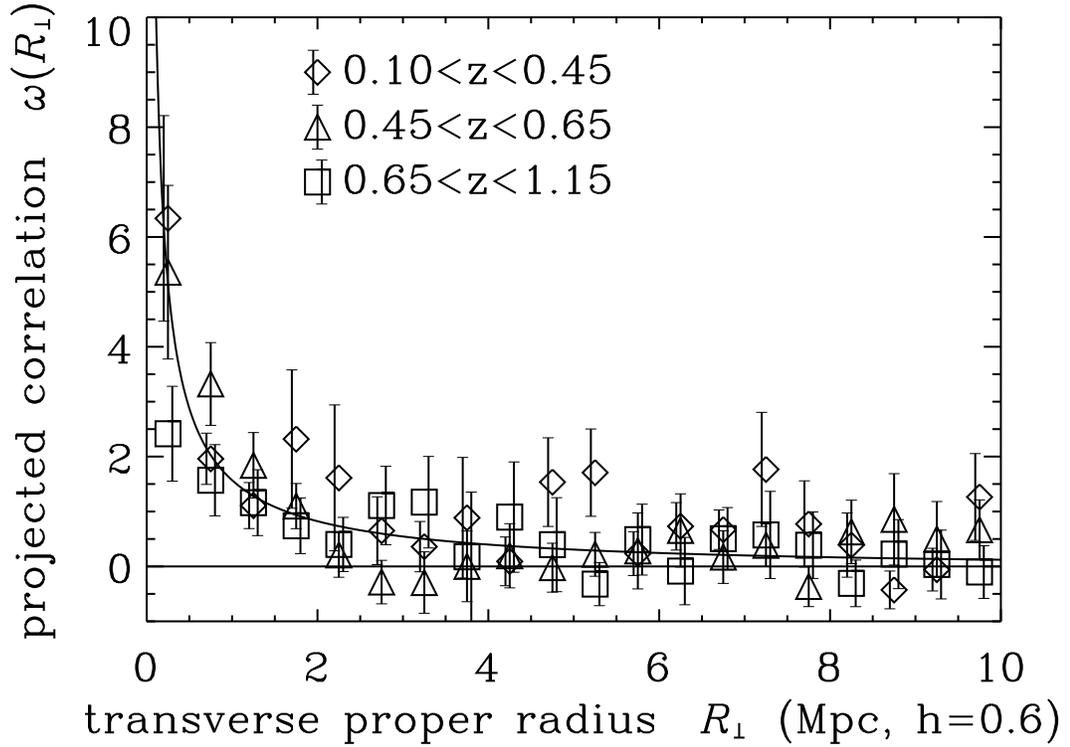}
\caption[]{The projected correlation function in proper coordinates in
the J0053+1234 region in three redshift bins.  The correlation function
is estimated as described in the text.  The solid curve shows a
correlation function of the form (\ref{eq:form}) with $r_0=3.5~{\rm
Mpc}$ and $\gamma=1.8$, projected through $\Delta\,v_r=1000~{\rm
km\,s^{-1}}$ at $z=0.55$ according to (\ref{eq:project}).}
\label{fig:wthetaz}
\end{figure}

\begin{figure}
\plotone{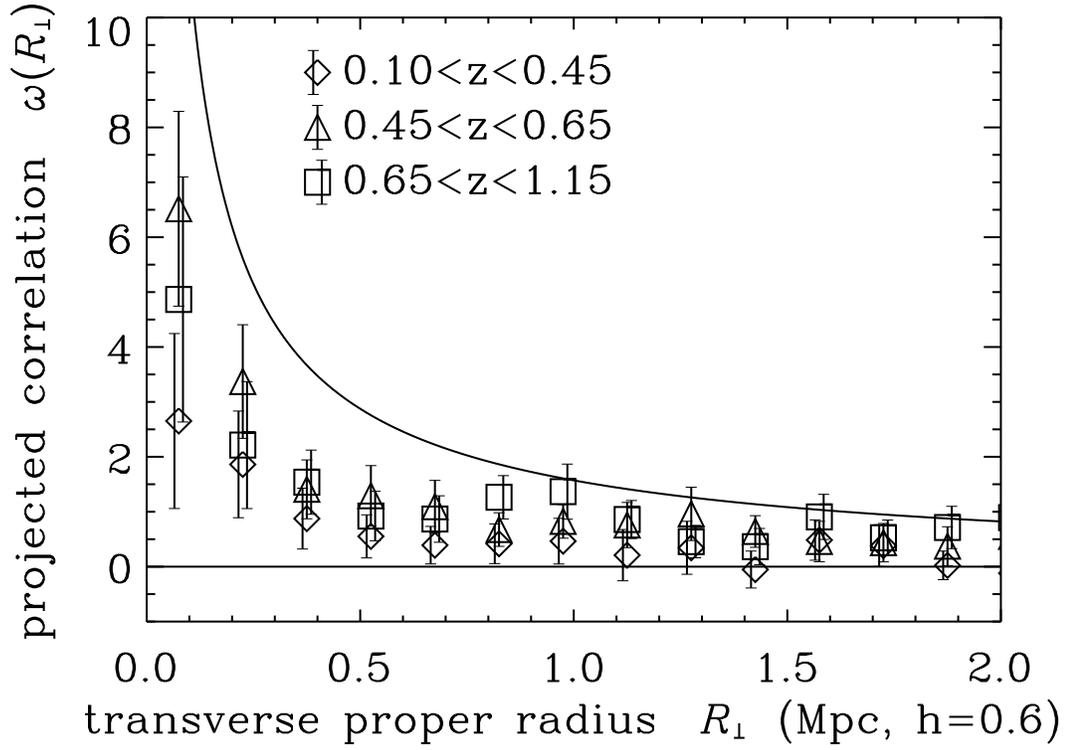}
\caption[]{The projected correlation function in proper coordinates in
the HDF region in three redshift bins.  The correlation function is
estimated as described in the text.  The solid curve shows the same
correlation function as that shown in Figure~\ref{fig:wthetaz}.  Note
that the horizontal axis scale is different from that in
Figure~\ref{fig:wthetaz} because the angular separation coverage is
smaller in the HDF sample than in the 0053+1234 sample.}
\label{fig:wtzhdf}
\end{figure}

\begin{figure}
\plotone{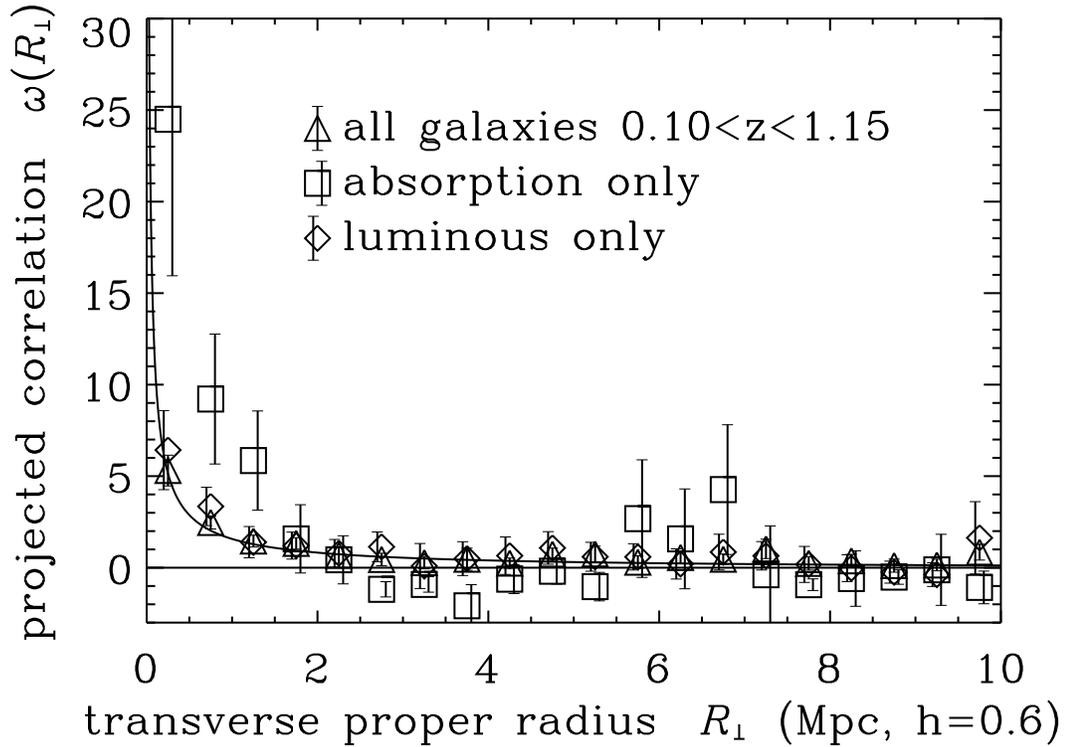}
\caption[]{The projected correlation function in proper coordinates
for the whole J0053+1234 sample, just those J0053+1234 galaxies with
absorption-line-dominated spectra, and just those J0053+1234 galaxies
with intrinsic luminosities (as estimated according to the text) $\log
L/L^{\ast}>-0.6$.  The solid curve shows a correlation function of the
form (\ref{eq:form}) with $r_0=3.5~{\rm Mpc}$ and $\gamma=1.8$,
projected through $\Delta\,v_r=1000~{\rm km\,s^{-1}}$ at $z=0.55$
according to (\ref{eq:project}).}
\label{fig:wthetat}
\end{figure}

The results of the correlation function estimation in the 0053+1234
region are shown in Figures~\ref{fig:wthetaz} and \ref{fig:wthetat}.
Results of the estimation in the HDF region are shown in
Figure~\ref{fig:wtzhdf}.  Error bars on the points are $1\,\sigma$,
computed by bootstrap resampling the D catalog.  It is important to
note that these bootstrap-resampling error bars do not include
uncertainties due to the specific R sample used (though this makes
only a tiny contribution to the overall error budget) nor
uncertainties due to sample variance, which, with a sample this small,
could be quite large.  Only intercomparison of similarly surveyed
fields will provide an empirical estimate of the uncertainties due to
sample variance.  On the other hand, the bootstrap errors may be
expected to overestimate the Poisson variance in measures of the
correlation function like those shown in Figures~\ref{fig:wthetaz} and
\ref{fig:wthetat} (Mo et al 1992).  The error bars (and their
covariances) are discussed at more length below.

Figures~\ref{fig:wthetaz} and \ref{fig:wtzhdf} show the projected
correlation function $\omega(R_{\perp})$ for the two samples split
into three redshift intervals.  These figures show that there is no
evidence for strong evolution in the proper correlation length with
redshift, although the proper correlation length is smaller in the
highest redshift subsample of the 0053+1234 sample, and smaller in the
lowest redshift subsample of the HDF sample.  The Figures also show a
theoretical correlation function which is the appropriate
line-of-sight projection of
\begin{equation}
\xi(r)=\left(\frac{r}{r_0}\right)^{-\gamma}
\label{eq:form}
\end{equation}
where the correlation length $r_0$ is set to $3.5$~proper Mpc with
$h=0.6$ and the exponent $\gamma$ is set to $1.8$.

\begin{figure}
\plotone{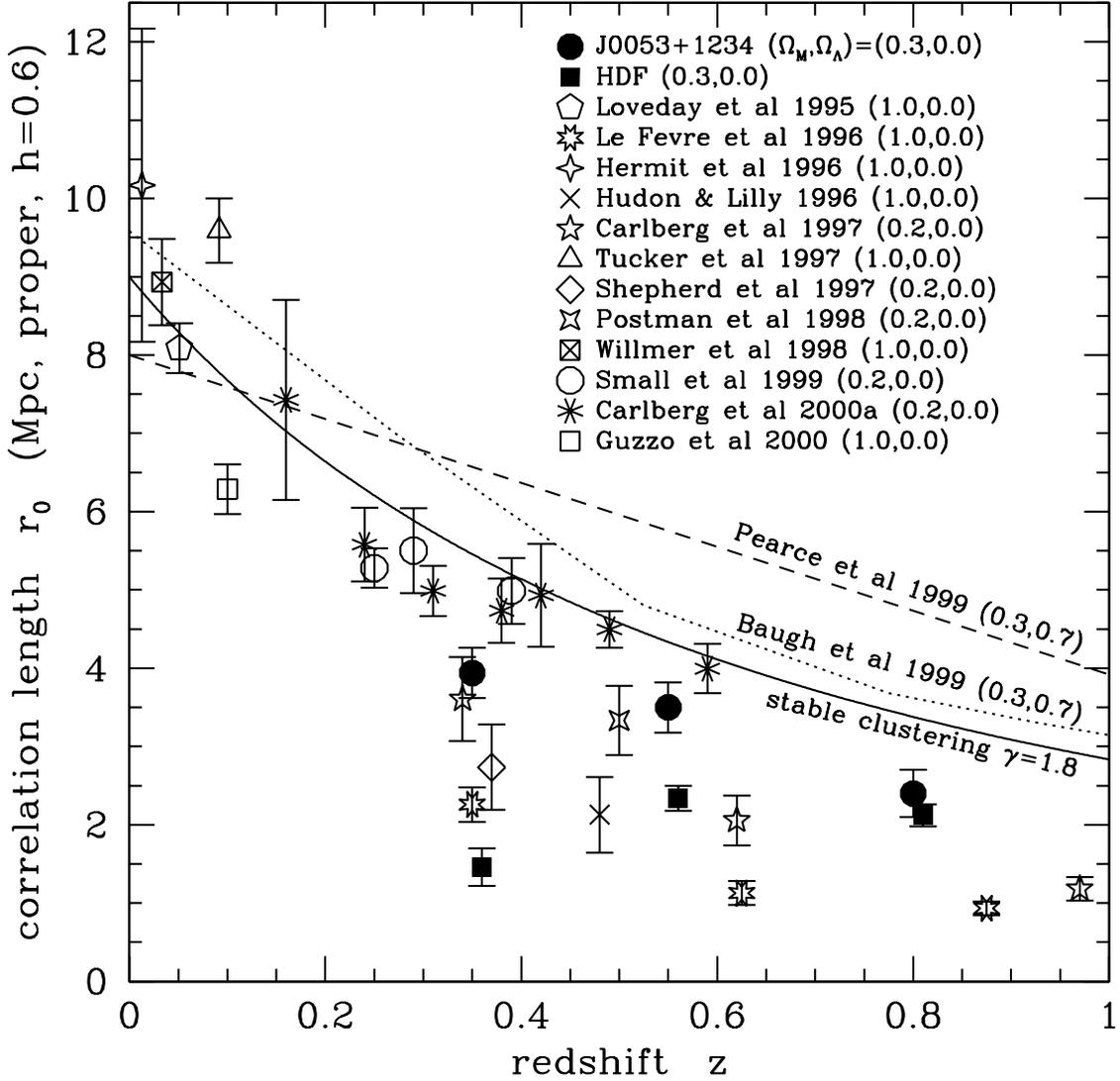}
\caption[]{The best-fit correlation lengths $r_0$ in this work and
from other studies.  For this work (solid points), the fits were
performed with $\gamma$ fixed at 1.8.  The solid line shows the
``stable clustering'' prediction (eg, Peebles 1993) for $\gamma=1.8$
with $r_0=9.0$~Mpc at $z=0$.  The results of the other work have been
transformed to $h=0.6$ and proper coordinates.  Error bars on the
solid points are based only on bootstrap resampling and therefore
represent lower limits to the true uncertainties.  The error bars on
the Le~Fe\'vre et al (1996) points are likely incorrect; see text.  It
should be noted that the other work is heterogeneous, with galaxies
selected in many different ways, in different magnitude and redshift
ranges, with different assumed or fit values of $\gamma$, and in
different cosmographic models.  See the original references for
details.  The dotted and dashed lines show theoretical predictions.}
\label{fig:others}
\end{figure}

The HDF sample shows smaller correlation lengths (weaker clustering)
than the 0053+1234 sample.  This could be due in part to the fact that
the HDF region was selected to be ``empty''; i.e., devoid of large,
bright galaxies, quasars, or radio sources (Williams et al 1996).
Since this clustering analysis probes the non-linear regime, such
selection criteria can strongly bias the results.  This explanation is
supported by the much lower clustering amplitude in the lowest
redshift bin (Figures~\ref{fig:wtzhdf} and \ref{fig:others}), which is
the bin most affected by avoidance of bright galaxies.  It has been
shown in models of structure formation that underdense regions of the
Universe show a lower overall clustering amplitude, although the shape
(ie, exponent) of the correlation function may be more universal (eg,
Scoccimarro et al 1999).

Figure~\ref{fig:wthetat} indicates that when the galaxies are
separated by spectral type (Cohen et al 1999a), the
absorption-line-dominated galaxies show stronger correlations at small
separations, as is seen in the local universe (eg, Davis \& Geller
1976; Hermit et al 1996; Willmer et al 1998).  Of course the sample of
absorption galaxies is small (only 121), so sample variance may make
this result somewhat uncertain.  One indication that this may be
important is that the points are not well-fit by a power-law
correlation function.  Figure~\ref{fig:wthetat} also shows that there
is not a very strong dependence of the clustering on absolute galaxy
luminosity.  For the purposes of this test, galaxy luminosities were
estimated in the rest-frame $R$ band, using the redshift and
k-corrections from the literature (Poggianti 1997).  A galaxy was
classified as having ``high luminosity'' if its rest-frame $R$-band
luminosity was estimated to brighter than $L^{\ast}/4$ where
$L^{\ast}$ is taken from the evolving models of Poggianti (1997).  The
fact that the luminosity trend found in the local Universe (eg,
Willmer et al 1998) was not confirmed may be a consequence of the
small sample size.

Correlation function models of the form (\ref{eq:form}) with
$\gamma=1.8$ fixed were fit to the points in Figures~\ref{fig:wthetaz}
and \ref{fig:wtzhdf} by least-squares, using the bootstrap
uncertainties.  The best fits and uncertainties, computed by the
one-parameter condition $\chi^2<\chi^2_{\rm best}+1$ where
$\chi^2_{\rm best}$ is the best-fit value of $\chi^2$ (Press et al
1992), are plotted in Figure~\ref{fig:others}.  The figure also shows
that the results presented here are consistent with other studies.

Figure~\ref{fig:others} contains a line which illustrates the
prediction of ``stable clustering,'' the best no-evolution model for
small-scale clustering of galaxies in which virialized groupings of
different scales separate without growing in the expanding Universe.
If the small-scale clustering of galaxies is basically accounted for
by the presence of groups whose proper number densities and radii do
not evolve with time, then the clustering length will decrease with
increasing radius as the mean cosmic density rises (because the
clustering radius is defined to be the radius at which the overdensity
is unity).  For a correlation function of the form (\ref{eq:form}),
clustered density falls with radius as $r^{-\gamma}$ and cosmic mean
rises with redshift as $(1+z)^3$, so the stable clustering prediction
is that the proper correlation length decreases with radius as
$(1+z)^{(-3/\gamma)}$.

More detailed theoretical models, based on extensions of the cold dark
matter hypothesis to include the relationship between observable
galaxy clustering and mass clustering, make predictions which are
remarkably similar to those of stable clustering (Baugh et al 1999;
Pearce et al 1999).  These models are also shown in
Figure~\ref{fig:others}.

All the studies of galaxies at redshifts $z>0.2$ show smaller
correlation lengths than expected based on the surveys of the local
Universe and the assumption of stable clustering.  This evolution
result should be treated with caution, because galaxies observed in
the $z>0.2$ samples have been selected and photometered with data of
different qualities and spatial resolutions, observed in different
photometric bandpasses, and collected in samples with very different
field areas and numbers of sources.  Any individual study, taken by
itself, including the present study, is at least marginally consistent
with stable clustering.

\subsection{Error bars}

It is very important to note that the results from the two sky regions
studied in this work differ by significantly more than their
shot-noise (as estimated by bootstrap resampling) error bars.  Of
course the field-to-field variations give a much more secure estimate
of the true uncertainties than bootstrap resampling in individual sky
regions.

One difficult point of comparison with other studies is in the size of
the error bars; there is no agreement about how such error bars should
be measured.  In particular, some (e.g., Carlberg et al 1997) divide
bootstrap error bars by $\sqrt{3}$ based on an analysis of the
statistical properties of bootstrap resampling (Mo et al 1992).  This
correction is not made here.  Some base error bars on internal
field-to-field variations.  One such study, the CFRS (Le~F\'evre et al
1996), shows error bars significantly smaller than the field-to-field
variations observed in the present work, despite having smaller
numbers and similar field sizes.  The field-to-field variations
observed here agree with those found in the CNOC survey (Carlberg et
al 2000a).

Very few surveys, if any, consider the covariances between the error
bars on separate points of the angular correlation function, even
though it is clear from the analysis techniques that such covariances
must be strong.  Fortunately, with the bootstrap technique, it is
possible to measure these covariances.  The covariance matrix (in the
form of correlation coefficients $r_{ij}$ defined by
\begin{equation}
r_{ij}\equiv\frac{\sigma_{ij}}{\sigma_i\,\sigma_j}
\end{equation}
where $\sigma_{ij}$ is the covariance between data points $i$ and $j$
and $\sigma_i$ is the square root of the variance of data point $i$)
for the first ten ``whole sample'' points in Figure~\ref{fig:wthetat}
is
\begin{equation}
\tiny
\left[\begin{array}{rrrrr rrrrr}
 1.00& 0.28& 0.18& 0.26&-0.11& 0.03& 0.11&-0.13&-0.10& 0.07\\
 0.28& 1.00& 0.14& 0.13&-0.30& 0.29& 0.03&-0.52&-0.22&-0.03\\
 0.18& 0.14& 1.00& 0.12&-0.11&-0.04& 0.05&-0.06& 0.32&-0.25\\
 0.26& 0.13& 0.12& 1.00&-0.09& 0.04& 0.22& 0.04& 0.00&-0.13\\
-0.11&-0.30&-0.11&-0.09& 1.00&-0.01& 0.04& 0.11& 0.11& 0.35\\
 0.03& 0.29&-0.04& 0.04&-0.01& 1.00& 0.30&-0.15&-0.12&-0.05\\
 0.11& 0.03& 0.05& 0.22& 0.04& 0.30& 1.00& 0.46& 0.13& 0.15\\
-0.13&-0.52&-0.06& 0.04& 0.11&-0.15& 0.46& 1.00& 0.23&-0.09\\
-0.10&-0.22& 0.32& 0.00& 0.11&-0.12& 0.13& 0.23& 1.00& 0.07\\
 0.07&-0.03&-0.25&-0.13& 0.35&-0.05& 0.15&-0.09& 0.07& 1.00
\end{array}\right]
\end{equation}
It is clear that many of the covariances are large and ought to be
taken into account in estimating error regions.  Furthermore, this
matrix only represents covariances from shot noise, and not the even
more significant covariances expected from the fact that the
correlation function grows by gravitational clustering (eg,
Scoccimarro et al 1999).  When $\chi^2$ is computed using the entire
covariance matrix rather than just the individual data-point
variances, the preferred ranges on the correlation length $r_0$
actually go down (i.e., improve) by tens of percent.  This statement
is true for this study, but will not be true in general, because the
values and signs of the covariances depend on the survey size and
geometry, along with the analysis technique.  In many cases, inclusion
of the covariances will loosen constraints on $r_0$.  The fact that
the covariances affect the confidence intervals, however, demonstrates
that such covariances should be taken into account in future studies.

\section{Pairwise velocity dispersion}
\label{sec:pairwise}

The pairwise velocity dispersion (roughly the average radial velocity
difference between pairs of nearby galaxies) is a combined measure of
the mean mass density of the Universe and the amplitude of the power
spectrum on relatively small scales (eg, Davis et al 1985).  Local
measurements vary somewhat by survey and technique but are generally
in the range $300$ to $600~{\rm km\,s^{-1}}$ (Davis \& Peebles, 1983;
Marzke et al 1995; Somerville et al 1997b; Landy et al 1998).

In the previous Section, the correlation function of galaxies was
estimated from line-of-sight projections which are insensitive to
line-of-sight velocities.  However, just as the transverse correlation
function can be measured, so can a line-of-sight velocity correlation
function, which is a convolution of the real-space two-point
correlation function and the pairwise velocity distribution of
galaxies.  The procedure is to compute a two-dimensional correlation
function, as a function of both transverse separation and
line-of-sight velocity difference, and then fit to models of the
correlation function consisting of a radial power-law convolved with a
line-of-sight velocity dispersion.  The estimator is the same as that
used to measure the transverse one-dimensional correlation function,
but applied to two-dimensional maps of the numbers of pairs in the D
and R catalogs.  The fits to the two-dimensional correlation function
provide similar correlation lengths to those derived from the fits to
the one-dimensional function, although with larger uncertainties.

The best-fit velocity dispersions are found to vary strongly from
field to field and from redshift interval to redshift interval, over
the range $200<\sigma_v<600~{\rm km\,s^{-1}}$.  The reason for this
variation is that although the prominent redshift features or groups
found in the pencil-beam samples contain only about half of the
galaxies, they contain the vast majority of the velocity-space galaxy
pairs.  This means that any measurement of pairwise velocity
dispersion will be dominated by the particular galaxy groups found in
that region, in that redshift range.  The pairwise velocity dispersion
is a statistic strongly dominated by the highest density regions of a
survey.  This is not a new result; indeed, local measurements of this
velocity dispersion in surveys of $10^3$ to $10^4$ galaxies are
strongly affected by the inclusion or exclusion of the richest groups
or clusters (eg, Marzke et al 1995, Somerville et al 1997b).  This is
also true of the simulations (Somerville et al 1997a).  Surveys
including several or many $10^4$ galaxies may be large enough to give
stable results (Landy et al 1998).

\section{Galaxy structures}
\label{sec:groups}

\begin{figure}
\plotone{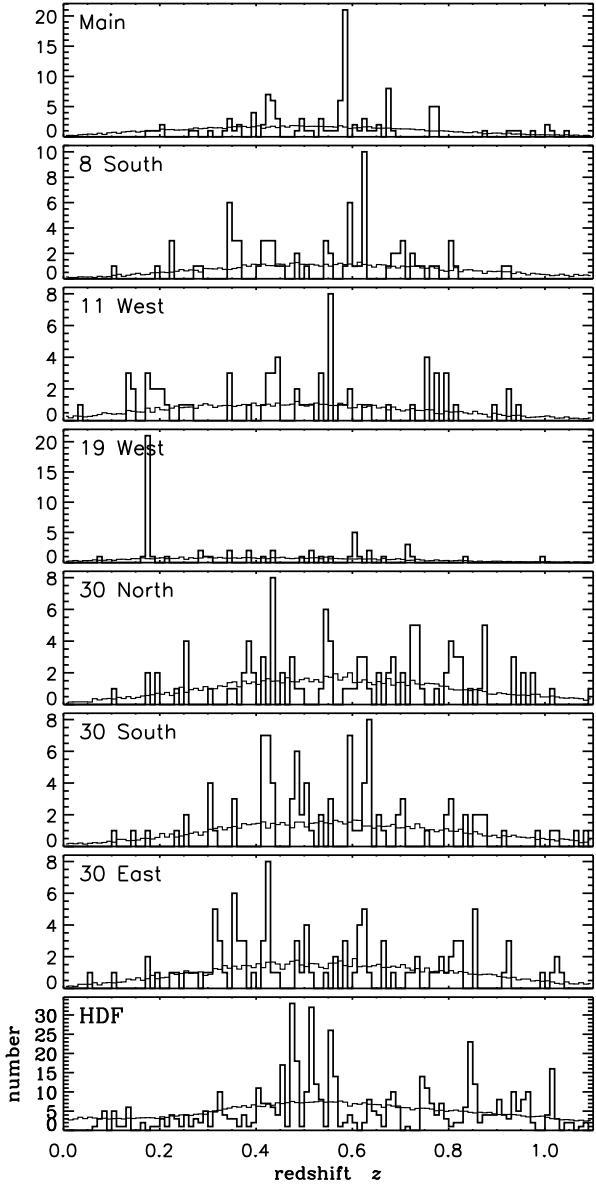}
\caption[]{The dark histograms show the numbers of sources as a
function of redshift in the subfields of the J0053+1234 region and in
the HDF region.  The light histogram shows the numbers in the random
catalog described in the text.  Because the R catalog is 100 times
larger than the data catalog, the R numbers have been divided by 100.
Note that different panels have different vertical scales.}
\label{fig:zhist}
\end{figure}

The redshift histograms for the fields are shown in
Figure~\ref{fig:zhist}, along with the radial velocity selection
functions implied by the random catalogs.  As has been noted
previously (eg, Broadhurst et al 1990; Crampton et al 1995; Cohen et
al 1996a; Cohen et al 1996b; Koo et al 1996), these ``pencil-beam''
redshift samples exhibit extremely strong clustering along the line of
sight in redshift space.  In the main subfield in the J0053+1234
region, with its complete sample, we estimated (Cohen etal 1996a) that
more than half of the galaxies lie in the five strongest redshift
features, which, except for one with a very high velocity dispersion,
probably correspond to poor groups of galaxies.  The spatial
distributions of the galaxies in these redshift features support this
(Cohen et al 1996a); they are non-uniform with little central
concentration (except in the strongest feature at $z=0.58$).  Similar
conclusions were reached in the HDF region (Cohen et al 1996b).  These
redshift features are well-sampled to $z=0.9$ and appear to persist
to higher redshifts.  At all redshifts, galaxies are often found in
pairs, presumably merging (eg, Carlberg et al 2000b).  It is apparent
from Figure~\ref{fig:zhist} that all the subfields show these redshift
features.

Five of the redshift features may correspond to Abell richness class
$-1$ or greater galaxy clusters, ie, clusters containing more than 20
galaxy members with absolute magnitudes brighter than two mag fainter
than the absolute magnitude of the third brightest member.  These
include the $z=0.58$ feature in the J0053+1234 main subfield, the
$z=0.175$ feature in the J0053+1234 19 West subfield, and three
features at $z=0.516$, $0.560$, and $0.848$ in the HDF region.  These
structures have velocity dispersions above $500~{\rm km\,s^{-1}}$ and
tens of galaxies within the brightest few magnitudes.  The presence of
these five rich structures in our combined samples seems reasonable;
an extrapolation to $z=0.8$ of the areal density of the sparsest
clusters in the Palomar Distant Cluster Survey (Postman et al 1996)
predicts that we should find some of these true clusters.

The statistics of these redshift-space features are discussed
elsewhere (Cohen et al 1996a, 1996b, 2000).  Briefly, they tend to
have velocity dispersions of $\sim 300~{\rm km\,s^{-1}}$, with large
uncertainties since they approach the redshift measurement
uncertainties of $\sim 100~{\rm km\,s^{-1}}$ (in line-of-sight
rest-frame velocity).  The galaxies in the features are spread across
the pencil-beam subfields, each of which spans a few proper Mpc at the
redshifts of interest.  Typical comoving line-of-sight distances
between adjacent features in a single subfield is $\sim 100~{\rm
Mpc}$.  The distribution of comoving radial separations between the
redshift-space features shows little evidence of a feature near
$100\,h^{-1}$~Mpc and no evidence for periodic structures seen in
previous studies (Broadhurst et al 1990).

The transverse size of the individual subfields is $8'\equiv3\pm0.5$
proper Mpc over most of the redshift interval of interest. We
introduce a proper longitudinal coordinate $Z\equiv ct(z)$ where
$t(z)$ is the age of the universe and just deal with structures whose
longitudinal extent in $Z$ exceeds $\sim3$ proper Mpc
($\equiv250-350$~km s$^{-1}$ over redshifts of interest).  We are also
interested in features whose longitudinal extent is less than the
transverse separation between subfields which is roughly $\sim10$
proper Mpc. Our procedure ({\it cf} Cohen {\it et al} 1998) is to
examine the overdensities after rebinning the data with all widths
$w_Z$ in the range 3-10~proper Mpc and all phases and to minimize the
probability of a feature arising by chance relative to a Poissonian
distribution. This gives us a procedure, albeit somewhat arbitrary,
for identifying and rank ordering features on the basis of their
relative significance.  This also allows us to include features and
sub-features together.  The rank-ordered (by statistical significance)
features, obtained by this procedure, with the number $N$ of galaxy
members and velocity dispersions $\sigma_v$, are given in
Table~\ref{tab:features}.  We have chosen only the eight most
significant features for the investigations of size and morphology
which follow.

\section{Galaxy structure size and morphology}
\label{sec:groupgal}

As the seven pencils in the J0053+1234 region span nearly a degree on
the sky, we can investigate the degree of clustering on larger linear
scales, $\sim25$ proper Mpc, than probed by the two-point correlation
function.  Previously (Cohen et al 1998) we have hypothesized that
structures on even larger scales ($\sim 100$~Mpc) are present in the
distribution of galaxies when the universe was less than half its
present age that resemble the giant voids and walls reported in the
local universe (eg, J\^oeveer et al 1978; Geller \& Huchra 1989;
da~Costa et al 1994; Shectman et al 1996; Vettolani et al 1997).  We
now have enough data to examine this hypothesis.

Much of the controversy that surrounds the description of clustering
on these scales stems from inadequate, quantitative descriptions of
large scale structure. When presented with data in which at most a few
large features are present, statistical measures like the two point
correlation function that assume random phases, or at least a huge,
fair sample of the Universe, are likely to be misleading.
Furthermore, genuinely random density fields typically exhibit
structures which appear, subjectively, to be non-random. Faced with
these difficulties, we shall limit ourselves to a description of the
structure that we have observed and a comparison with the structure
that we find in control locations.  (We have attempted to perform more
quantitative analyses based upon the Karhunen-Loeve transformation and
the ``counts in cells'' method but find that these do not give stable
answers with these samples.)

\begin{figure}
\plotone{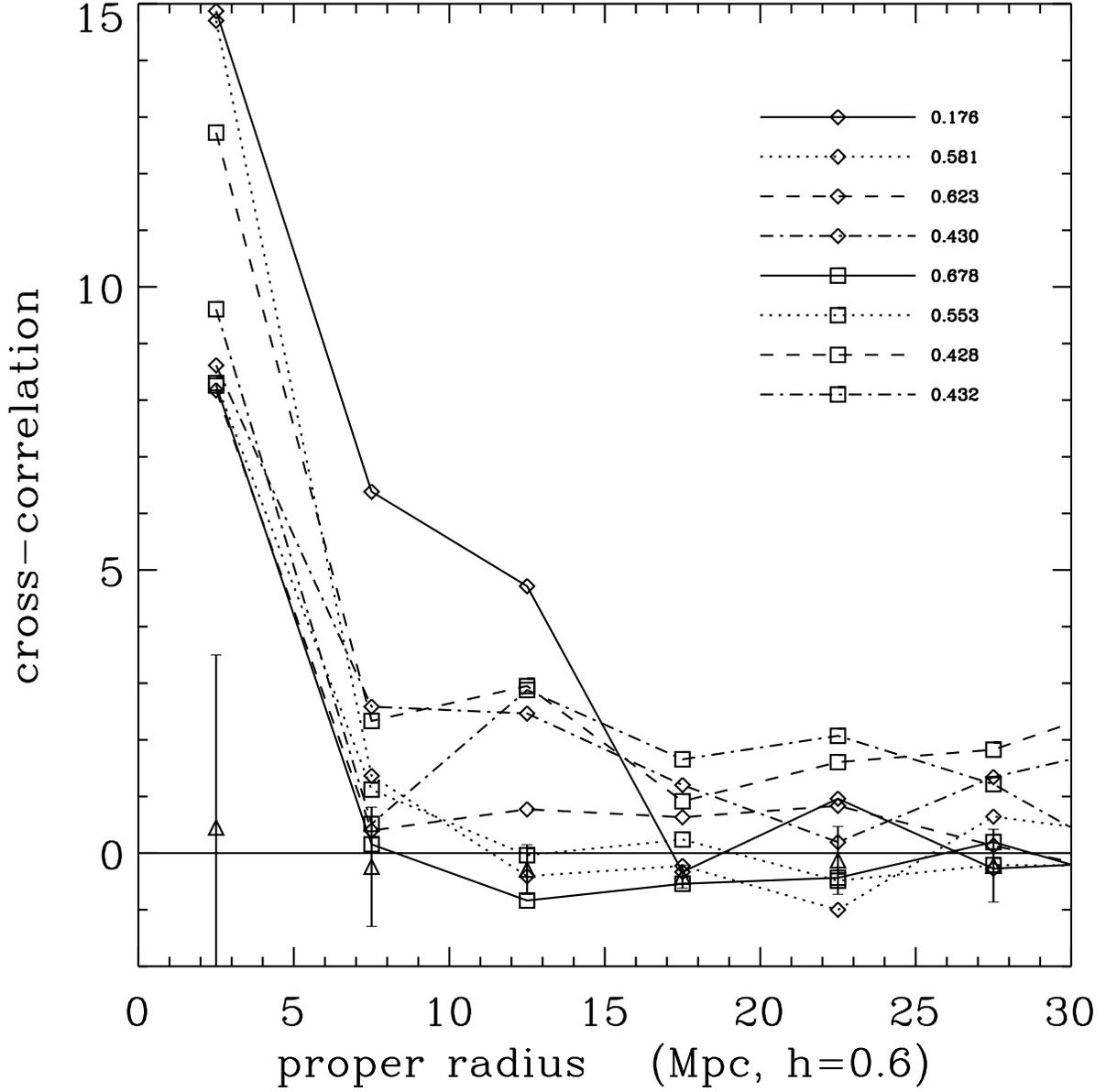}
\caption[]{The eight group-galaxy cross-correlation functions for the
eight redshift features in Table~\ref{tab:features} and the averaged
cross-correlation functions for the shifted control features
(triangles).  The error bars represent the control feature-to-feature
scatter.  Details in the text.}
\label{fig:groupgal}
\end{figure}

Three of the most significant redshift features are in the vicinity of
$z=0.430$.  These three features lie along the north-south axis
passing through the Main subfield and they span the full degree of the
J0053+1234 region.  This suggests large galaxy structures that span
several subfields.  In order to test the hypothesis that such large
structure is generic, each of the eight identified features (in
Table~\ref{tab:features}) was taken in turn and the cross-correlation
with individual galaxies out to large proper radii was computed.
Specifically, all galaxies located within proper distances $R<30$ Mpc
of the group center were binned.  These numbers were ratioed with the
same for the random catalog and unity was subtracted to make a
standard cross-correlation function.  For comparison with a control
sample, the exercise was repeated with the same group centers
displaced in redshift by 0.1.  The results are shown in
Figure~\ref{fig:groupgal}, with error bars on the control points found
by taking the scatter among the results for each of the eight control
redshift features.  No individual group shows a regular, power-law
cross-correlation, so it is not possible to objectively fit the
individual functions.  Roughly speaking, cross-correlation lengths of
5 to 20~proper Mpc are found, with large group-to-group scatter.
These sizes, combined with the measured velocity dispersions of a few
$100~{\rm km\,s^{-1}}$ imply virial masses from a few $10^{13}$ to a
few $10^{14}\,M_{\odot}$.  Crossing times are several $10^9$ years.
The groups near $z=0.430$ show the largest correlation lengths because
there are three of them, spanning $\sim 1$~deg on the sky.  Galaxy
structures of this size have been found previously (Small et al 1999).

There is a second hypothesis that can be tested with this data
set. This is that galaxies are concentrated on planes, similar to
those that have been reported locally and are sometimes found in
numerical simulations. In order to define a plane, we must consider
three non-colinear points.  For each feature (group center), any two
additional galaxies outside the feature define, along with the feature
center a plane.  The unit normal vector (3-vector) to this plane can
be constructed.  The direction cosines of this normal vector are
defined to be the two components of the vector transverse to the line
of sight; the projection of the normal vector onto the plane of the
sky.  (Those galaxies which lie in the same subfield as the feature
and those pairs for which the subfields are colinear with the feature,
for which the planes are poorly defined, are excluded.)

\begin{figure}
\plotone{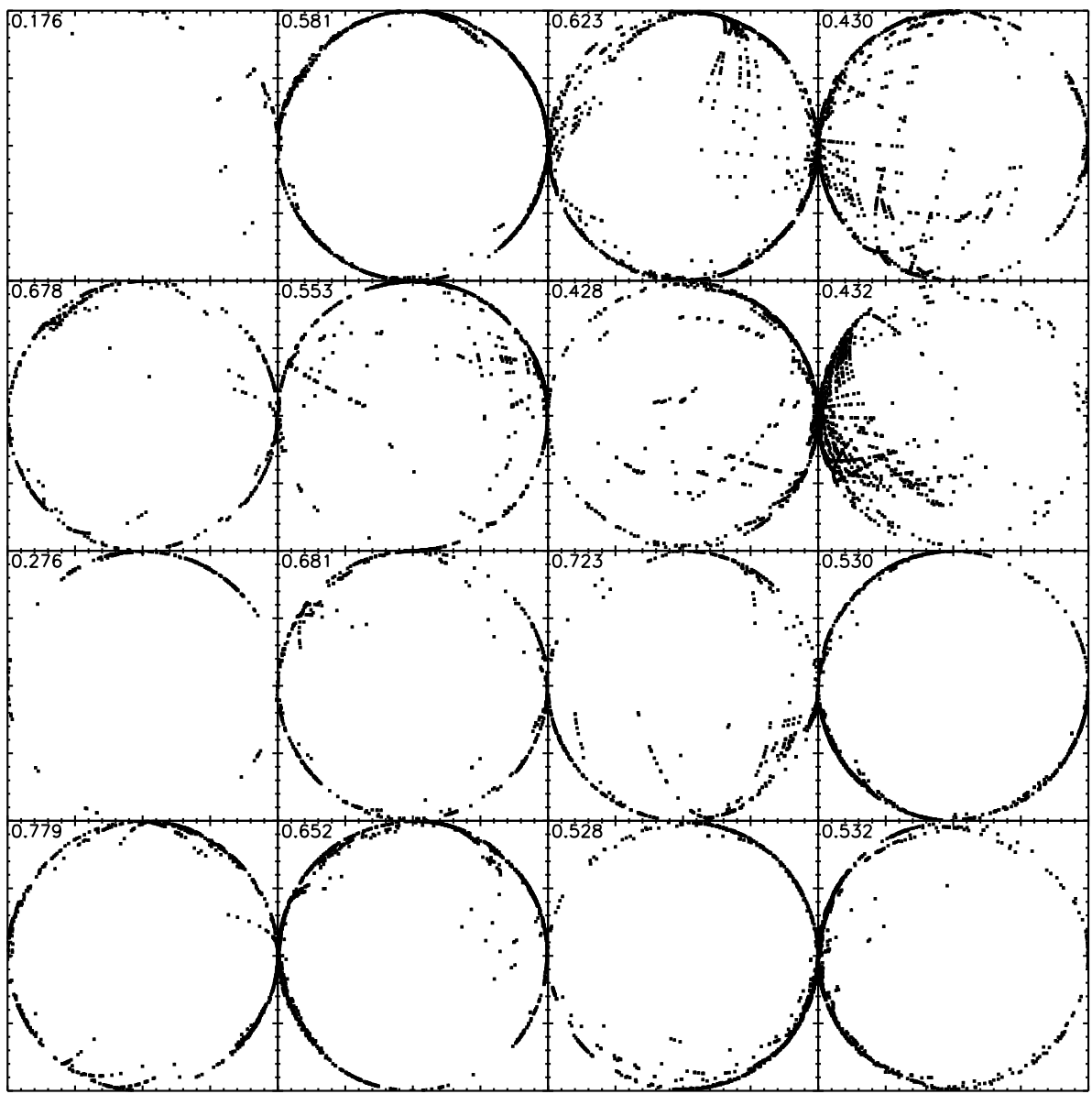}
\caption[]{The direction cosines for pairs of galaxies relative to
each of the eight groups in Table~\ref{tab:features} (top eight
panels) and for the same group centers but shifted in redshift by 0.1
(bottom eight panels).  Each point in each panel shows the
north--south (vertical) and east--west (horizontal) components of the
unit 3-vector normal to the plane defined by each triplet composed of
each pair of galaxies and the group center.  Details in the text.}
\label{fig:dircos}
\end{figure}

The direction cosines for each triplet (two galaxies and the feature
center) are plotted on a plane, for each feature and a set of
controls, in Figure~\ref{fig:dircos}. If the features lie in prominent
planes of galaxies, these will show up as clustering of points on the
direction cosine plane.  As usual, the controls were made by
translating the features in redshift by 0.1.  There is no clear
evidence that galaxies are concentrated on planes from these data.

It is notable, however, that three of the highly significant redshift
features (near $z=0.43$) ar roughly colinear.  Figure~\ref{fig:spine}
shows the galaxy configuration in the region of these galaxy
structures.  They may represent part of a filament, a structure
generic to the numerical simulations.  Clearly studies such as these
of the morphologies of large-scale structures will benefit from large,
densely sampled, contiguous areas of sky.  The sparse coverage of the
J0053+1234 field, though necessary for probing large angular scales,
is not sufficient for making definitive statements about structure
morphology.

\section{Summary}
\label{sec:summary}

A clustering analysis is performed on two samples of $\sim 600$ faint
galaxies each, in two widely separated regions of the sky.  One of the
survey regions is configured so that there are galaxy pairs spanning
angular separations of up to $1$~deg.  The median redshifts of the
galaxy samples are both $z_{\rm med}\approx 0.55$.

No strong evolution of the proper galaxy--galaxy correlation length
with redshift is found (relative to stable clustering), although
samples of $\sim 1000$ sources are simply not large enough for
detecting subtle changes.  There is some evidence for
weaker-than-stable clustering at intermediate redshift, when the
results presented here are compared with surveys of the local
($z<0.2$) Universe, but it is to be cautioned that this result depends
on comparing samples selected and observed with very different
techniques.  Strong field-to-field variations are found, possibly
emphasized by the stringent ``blank-field'' criteria used to select
the HDF.

Galaxies with absorption-line-dominated spectra show much stronger
clustering at proper distances of $<2$~Mpc than typical field
galaxies, in qualitative similarity to the morphological segregation
observed in the local Universe.

Although pairwise velocity dispersions are measured to be in the range
$200$ to $600~{\rm km\,s^{-1}}$, measured value for any particular sky
region and redshift interval is dominated by the prevalent individual
redshift features.  For this reason, no statement about the evolution
in pairwise velocity dispersion is possible from these data.

Each individual pencil-beam redshift sample contains of order five
significant features (``peaks'') in its redshift distribution to
redshift unity.  Associating the redshift features with physical
galaxy structures, we have performed a three-dimensional
structure--galaxy correlation analysis and find that galaxies show
measurable clustering around these structures out to $\sim 20$~Mpc.
These lengthscales are longer than the galaxy--galaxy correlation
length and similar to scales of superclusters found locally and,
indeed, at even higher redshifts (Lubin et al 2000).  There is some
evidence for filaments, although pencil-beam surveys are not ideally
suited for finding them in general.

The most prominent galaxy structures have velocity dispersions of a
few $100~{\rm km\,s^{-1}}$ and coherence lengths of 5 to 15 proper Mpc
in the adopted cosmology.  These physical parameters imply masses of
$\sim 3\times10^{13}$ to $3\times 10^{14}\,M_{\odot}$.  The galaxies
in the structures are not obviously concentrated in walls or sheets,
and no evidence is found for periodicity on very large length scales.
On the other hand, this survey of spatially separated, sparsely
sampled pencil beams is not extremely sensitive to all kinds of
structure morphology and will be outmoded by future surveys with
larger contiguous solid angle coverage.

Most importantly, large-scale structure at $z\sim 1$ does not look
very different from that found locally at the present epoch.

\acknowledgements It is a pleasure to thank Ray Carlberg, Daniel
Eisenstein, Simon Lilly, Jim Peebles, Roman Scoccimarro, Todd Small
and an anonymous referee for useful comments, discussions and
assistance with the literature.  This research is based on
observations made at the W. M. Keck Observatory, which is operated
jointly by the California Institute of Technology and the University
of California; and at the Palomar Observatory, which is owned and
operated by the California Institute of Technology.  Financial support
was provided under NSF grant AST~95-29170 (to RDB) and Hubble
Fellowship grant HF-01093.01-97A (to DWH) from STScI, which is
operated by AURA under NASA contract NAS~5-26555.  This research made
use of the NASA ADS Abstract Service.

\begin{deluxetable}{ccccccc}
\tablewidth{0pt}
\tablecaption{
	Subfields in the J0053+1234 region
	\label{tab:fields}
}
\tablehead{
   \colhead{subfield}
 & \colhead{$\Delta\alpha$}
 & \colhead{$\Delta\delta$}
 & \colhead{$\Omega$\tablenotemark{a}}
 & \colhead{flux range}
 & \colhead{$N_z$\tablenotemark{b}}
 & \colhead{$f_z$\tablenotemark{c}}
\\
 & \colhead{(arcmin)}
 & \colhead{(arcmin)}
 & \colhead{(arcmin$^2$)}
 & \colhead{(mag)}
}
\startdata
main     &   $0.0$ &   $0.0$ & 15 &      $K<20.0$ & 139 & 0.84 \nl
8 South  &  $+4.6$ &  $-9.4$ & 14 & $22.0<R<23.5$ &  96 & 0.93 \nl
11 West  & $-11.7$ &  $-7.9$ & 14 & $22.0<R<23.5$ &  78 & 0.76 \nl
19 West  & $-19.3$ &  $-3.2$ & 14 & $22.0<R<23.5$ &  62 & 0.60 \nl
30 North &  $-0.6$ & $+30.5$ & 48 & $22.0<R<23.5$ & 119 & 0.34 \nl
30 South &  $-0.3$ & $-32.7$ & 47 & $22.0<R<23.5$ & 116 & 0.33 \nl
30 East  & $+29.5$ &  $-0.1$ & 48 & $22.0<R<23.5$ & 119 & 0.34 \nl
\hline
total    &         &         &    &               & 729 \nl
\enddata
\tablenotetext{a}{$\Omega$ is the solid angle of each field.}
\tablenotetext{b}{$N_z$ is the number of spectroscopically confirmed
galaxies with redshifts.  Spectroscopically confirmed stars have been
excluded.}
\tablenotetext{c}{$f_z$ is the approximate fraction of galaxies in the
magnitude range whose redshifts have been measured.}
\end{deluxetable}

\begin{deluxetable}{cccc}
\tablewidth{0pt}
\tablecaption{
	Highly significant redshift features in the J0053+1234 region
	\label{tab:features}
}
\tablehead{
   \colhead{redshift}
 & \colhead{$\sigma_v$\tablenotemark{a}}
 & \colhead{subfield}
 & \colhead{$N$}
\\
 & \colhead{$\rm (km\,s^{-1})$}
}
\startdata
0.173\tablenotemark{b} & 473 & 19 West  & 20 \nl
0.577\tablenotemark{c} & 404 & Main     & 24 \nl
0.621 & 278 & 8 South  & 10 \nl
0.428 & 346 & Main     & 10 \nl
0.676 & 298 & Main     &  9 \nl
0.550 & 274 & 11 West  &  8 \nl
0.427 & 255 & 30 South &  7 \nl
0.431 & 165 & 30 North &  7 \nl
\enddata
\tablenotetext{a}{No correction to the velocity dispersion $\sigma_v$ was
made for measurement uncertainties.}
\tablenotetext{b}{Combination of 3 significant sub-features.}
\tablenotetext{c}{Combination of 2 significant sub-features.}
\end{deluxetable}

\cleardoublepage

\begin{figure}
\caption[]{[NOTE TO astro-ph USER: These figures are in the archive as
rdb043.gif and rdb053.gif] {\sl (Top)} Location of galaxies in a
$30\times30\times100$~proper Mpc (equivalent to cosmic time along the
line of sight) region of the J0053+1234 field chosen to include three
distinct redshift features near $z=0.43$ in Table~\ref{tab:features}.
The faint ``ribs'' connecting individual galaxies to the central
``spine'' (chosen to lie at the center of the main field) are
perpendicular to the spine.  Our typical velocity error of $200$~km
s$^{-1}$ Mpc$^{-1}$ corresponds to a distance of 2.4 Mpc along the
line of sight and departures from Hubble expansion are less than
$\sim7$~Mpc.  The radius of a galaxy symbol corresponds to
$M_R^\ast-M_R$. The galaxies are color-coded with red corresponding to
absorption line galaxies, green composite and blue emission line
galaxies. Note that absorption line galaxies are more luminous and
concentrated in the redshift features. It is apparent that galaxies
are clustered around this concentration with correlation lengths of
$\sim20-30$~proper Mpc.  The three principal features are roughly
colinear, but, in general, there is little indication that galaxies
are arranged either on planes or lines.  {\sl (Bottom)} A more typical
region, with similar dimensions, displaced from that shown in the
first panel by $\Delta z=0.1$.}
\label{fig:spine}
\end{figure}

\end{document}